%% file: Swiss_system_efficacy_competitiveness_trade_off.tex
\documentclass[12pt]{article}
\usepackage[utf8]{inputenc}
\usepackage[british]{babel}
\usepackage{cmap}
\usepackage{lmodern}

\usepackage{amssymb, amsmath, amsthm}
\usepackage[a4paper,top=25mm,bottom=25mm,left=25mm,right=25mm]{geometry}
\usepackage{ragged2e}

\usepackage{authblk} 
\usepackage{pifont}
\usepackage{graphicx}
\usepackage[dvipsnames,svgnames,table]{xcolor}
\usepackage[figuresright]{rotating}
\usepackage{xtab} 
\usepackage{longtable} 
\usepackage{multirow}
\usepackage{footnote}
\usepackage[stable]{footmisc}
\usepackage{chngpage} 
\usepackage{pdflscape} 
\usepackage{tocbibind} 

\usepackage{pgfplots}
\pgfplotsset{compat=1.18}
\pgfplotsset{every tick label/.append style={font=\footnotesize}}
\usepackage{setspace}

\makesavenoteenv{tabular}
\usepackage{tabularx}
\usepackage{booktabs}
\usepackage{threeparttable} 
\usepackage[referable]{threeparttablex} 
\newcolumntype{R}{>{\raggedleft\arraybackslash}X}
\newcolumntype{L}{>{\raggedright\arraybackslash}X}
\newcolumntype{C}{>{\centering\arraybackslash}X}
\newcolumntype{A}{>{\columncolor{gray!25}}C}
\newcolumntype{a}{>{\columncolor{gray!25}}c}

\newlength{\tablen}

\usepackage{dcolumn} 
\newcolumntype{.}{D{.}{.}{-1}}

\usepackage{tikz}
\usetikzlibrary{arrows, calc, matrix, patterns, positioning, trees}
\usepackage[semicolon]{natbib}
\usepackage[hyphens]{url}
\usepackage{hyperref} 
\hypersetup{
  colorlinks   = true,    		
  urlcolor     = blue,    		
  linkcolor    = blue,			
  citecolor    = ForestGreen,		  	
}
\usepackage{microtype}
\usepackage[justification=centering]{caption} 

\usepackage[labelformat=simple]{subcaption}

\DeclareCaptionLabelFormat{parenthesis}{(#2)}
\makeatletter
\renewcommand\p@subfigure{\arabic{figure}.}
\makeatother

\DeclareCaptionLabelFormat{parenthesis}{(#2)}
\makeatletter
\renewcommand\p@subtable{\arabic{table}.}
\makeatother

\usepackage{alphalph}

\def\addlegendimage{\csname pgfplots@addlegendimage\endcsname}

\usepackage{enumitem}

\setlist[itemize]{leftmargin=2.5\parindent}
\setlist[enumerate]{leftmargin=2.5\parindent}

\theoremstyle{plain}

\theoremstyle{definition}

\theoremstyle{remark}

\def\keywords{\vspace{.5em} 
{\noindent \textit{Keywords}: }}

\def\AMS{\vspace{.5em} 
{\noindent \textbf{\emph{MSC} class}: }}

\def\JEL{\vspace{.5em} 
{\noindent \textbf{\emph{JEL} classification number}: }}

\title{The efficacy-competitiveness \\ frontier in the Swiss system}
\author{
\href{https://sites.google.com/view/laszlocsato}{L\'aszl\'o Csat\'o}\thanks{~Corresponding author \newline
E-mail: \emph{laszlo.csato@sztaki.hun-ren.hu} \newline
Institute for Computer Science and Control (SZTAKI), Hungarian Research Network (HUN-REN), Laboratory on Engineering and Management Intelligence, Research Group of Operations Research and Decision Systems, Budapest, Hungary \newline 
Corvinus University of Budapest (BCE), Institute of Operations and Decision Sciences, Department of Operations Research and Actuarial Sciences, Budapest, Hungary}
$\qquad \qquad$
Bal\'azs R.~Sziklai\thanks{~E-mail: \emph{sziklai.balazs@krtk.elte.hu} \newline
Institute of Economics, ELTE Centre for Economic and Regional Studies, Budapest, Hungary \newline
Corvinus University of Budapest (BCE), Institute of Operations and Decision Sciences, Department of Operations Research and Actuarial Sciences, Budapest, Hungary}
}
\date{\today}

\begin{document}

\newgeometry{top=25mm,bottom=25mm,left=25mm,right=25mm}
\maketitle
\thispagestyle{empty}

\begin{abstract}
\noindent
An ideal tournament is both efficacious and attractive: it selects the best player as the winner, but only at the end of the tournament in order to maintain suspense as long as possible. We identify and explore the trade-off between these criteria in Swiss-system chess team tournaments via simulations. Organising more rounds is shown to improve efficacy but increase the probability of a stakeless last round. The 2008 reform in the Chess Olympiad ranking is found to be a questionable move, as the previous rule based on board points determines the efficacy-competitiveness frontier, regardless of whether two or three points are awarded for winning a match. Combining board points and match points is better than using match points only. Our results can help optimise the design of the Swiss system, which, besides chess, is also widely used in e-sports.
\end{abstract}


\AMS{62F07, 68U20, 90-10, 90B90}

\JEL{C44, C53, Z20}

\clearpage
\restoregeometry

\section{Introduction} \label{Sec1}

According to a recent survey of tournament design, efficacy and attractiveness are among the most important design criteria \citep{DevriesereCsatoGoossens2025}. \emph{Efficacy} is the ability of the tournament to approxiamte the latent strength ranking of the players by the ranking based on the results of matches. Indeed, a tournament is often considered a selection mechanism used to identify the best team in a noisy environment \citep{Ryvkin2010}. While attractiveness has several different interpretations, they are all connected to the entertainment generated by the tournament. Here we adopt a straightforward definition called \emph{competitiveness}: the winner should not be decided before the last round of the tournament.
This choice is supported by empirical findings. For example, elevated interest is detected for football matches that are potentially significant for end-of-season outcomes \citep{BuraimoForrestMcHaleTena2022}, and search interest declines substantially in the Formula One World Drivers' Championship after the winner becomes known \citep{Garcia-del-BarrioReade2022}.
Avoiding a stakeless match played by the final winner is also important from the perspective of fairness, since its opponent can win more easily in the absence of incentives to win, as the favourite probably does not exert full effort \citep{CsatoGyimesi2026a}.

Even though efficacy and competitiveness are both crucial characteristics of tournaments, we do not know of any previous study that explores the trade-off between them. We address this research gap by investigating the Swiss system.
The Swiss system is a non-eliminating tournament format, where the contestants play a given number of rounds chosen by the decision-maker, which gives a crucial flexibility missing from knockout and round-robin tournaments.

Aside from the adaptable nature of the Swiss system, our focus on it is explained by three reasons.
First, according to an extensive simulation study by \citet{SziklaiBiroCsato2022}, the Swiss system provides the best trade-off between efficacy and the number of matches among the most popular tournament formats.
Second, the last round became uncompetitive in some Swiss-system chess tournaments. For instance, in the 2015 Australian Open, the Chinese Grandmaster \emph{Ni Hua} scored 9.5 points in the first 10 rounds, but the runners-up had only 7.5 points when one point was awarded for a win, making the last, 11th round uninteresting to most viewers (see \url{https://s2.chess-results.com/tnr157180.aspx}).
Third, the Chess Olympiad, one of the most prestigious chess tournaments, uses the Swiss system. Furthermore, its ranking rules were fundamentally changed in 2008: since then, the ranking is based on the number of match points instead of the number of board points.

The 2008 reform had two important consequences. First, it made the margin of victory irrelevant, except for tie-breaking rules: a narrow match victory and a large-margin victory have the same reward in the primary criterion of the ranking. Second, it introduced an additional strategic consideration for the teams. Chess games, especially at higher levels, often end in draws. Players can often steer a game towards a draw by simplifying the position and avoiding unnecessary risks. Even in an advantageous position, it may be rational to settle for a draw if this secures a favorable team result. The two most obvious case is if three boards have already produced decisive results, and a draw on the remaining board is sufficient either to secure a victory for the team or to guarantee a drawn match. Indeed, this behaviour is frequently observed in team competitions. For example, during the 2018 Chess Olympiad, the Armenian player \emph{Lilit Mkrtchian} was reportedly instructed by her captain to seek a draw against \emph{Irina Krush}, despite having winning chances (see \url{https://www.chess.com/news/view/chess-olympiad-u-s-joins-poland-azerbaijan-in-lead}).

This mechanism may further increase the number of drawn games.
One way to counteract this tendency could be providing stronger incentives for taking risks. Therefore, some team chess competitions, such as the Global Chess League, award three points for winning a match. There is an ongoing debate about how effective this rule is in competitions where each team acts as a single decision-making entity, such as in football \citep{LentenKendall2022}. It is even less clear whether it has the same effect in team chess, where the incentives of the team may differ from those of its individual players.


The decision to use match points was probably motivated by a desire to strengthen the team-competition nature of the tournament, while awarding extra points for match victories aims to encourage teams to play for a win. Both objectives are valid, but they can also affect efficacy and competitiveness as we will see.

The paper is organised as follows. Section~\ref{Sec2} gives an overview of previous works on similar problems. The simulation method is described in Section~\ref{Sec3}, and the results are presented in Section~\ref{Sec4}. Finally, Section~\ref{Sec5} summarises the key findings and discusses directions for future research.

\section{Related literature} \label{Sec2}

The Swiss system has been developed to rank a large number of players who play a relatively small number of rounds without elimination. The pairing is dynamic; it depends on the results of the previous rounds: players generally play against opponents with (roughly) equal running score, subject to various constraints.
This format, proposed by the Swiss teacher Dr.~Julius M\"uller, was applied first in 1895 at the 5.~Schweizerische Schachturnier, but is now used in other games such as bridge and Go. E-sports competitions---including Counter-Strike: Global Offensive (CS:GO), Dota 2, Hearthstone, and Rocket League---often use a variant of the Swiss system \citep{DongRibeiroXuZamoraMaJing2023}.

The two major design issues in the Swiss system are the choice of the pairing algorithm and the ranking method.
Pairing is traditionally done by hand, but today several software are available for this purpose; see, for example, the pioneering study \citet{Olafsson1990} and the program Swiss-Manager (\url{https://swiss-manager.at/}). There are alternative pairing algorithms in the literature: \citet{KujansuuLindbergMakinen1999} determine the pairs by solving a stable roommates problem, \citet{GlickmanJensen2005} consider a Bayesian optimal design framework for pairing, while \citet{SauerCsehLenzner2024} suggest a parametric method based on finding maximum weight matchings in an appropriately constructed graph.
The main challenge of ranking resides in the uncertain trade-off between the number of points scored and the strength of opponents \citep{Chebotarev1994, Csato2013a, Csato2017a, CsatoDevriesereGoossensGyimesiLambersSpieksma2026, LeivaBertran2025}. Tie-breaking rules also play an important role in the Swiss system due to the low number of rounds compared to the number of players, especially in the middle of the ranking. \citet{Freixas2022} uncovers some weaknesses of the widely used Buchholz method, the arithmetic mean of points scored by all opponents.

Efficacy, the ability of a tournament to rank the players according to their (unknown) strengths, is usually examined via Monte Carlo simulations as the observed set of match results is only one realisation of the interaction of several random variables.
If the match outcomes are deterministic, ranking in the Swiss system with eight players and three rounds is perfect for the first and the last positions, but not for the middle positions \citep{DinhBaoKhalidIida2020}.
\citet{Appleton1995} computes the chance that the best player wins with random tie-breaking, is the sole winner, or shares the first place without tie-breaking when 8 or 16 players compete in five rounds. 
\citet{SziklaiBiroCsato2022} determine the efficacy of the Swiss system for 32 players under six different models of winning probabilities and several metrics of ranking ability. Similar to the current study, the number of rounds is assumed to be a decision variable.

Our measure of competitiveness, the probability of a stakeless last round, has received serious attention in (incomplete) round-robin tournaments. \citet{ChaterArrondelGayantLaslier2021} consider stakeless matches in various round-robin formats of the FIFA World Cup group stage, and \citet{CsatoMolontayPinter2024} in double round-robin groups of four teams. \citet{CsatoGyimesi2026a} study indifferent teams in the last round of the 2026 FIFA World Cup group stage. \citet{Gyimesi2024}, \citet{CsatoGyimesi2026b}, and \citet{DevriesereGoossensSpieksma2026} compare the pre-2024 and post-2024 formats of the UEFA Champions League from this perspective. \citet{Csato2023a} explores the effect of tie-breaking rules on the competitiveness of matches played in the last round of the four groups in the 2022/23 UEFA Nations League A. Finally, \citet{FeddersenHumphreysSoebbing2023} document that betting markets believe European football teams adjust their effort level according to the incentives at the end of the season; thus, the impact of incentives can actually be seen in
historical data.

\section{Methodology} \label{Sec3}

Following the standard approach in tournament design \citep{DevriesereCsatoGoossens2025}, we use Monte Carlo simulations to quantify efficacy and competitiveness.
Section~\ref{Sec31} explains the choice of team strengths. Section~\ref{Sec32} describes the pairing mechanism and the probabilistic model of \citet{Milvang2015} used to generate match results. The tournament metrics are presented in Section~\ref{Sec33}, and the details of the computations are discussed in Section~\ref{Sec34}.

\subsection{Input data} \label{Sec31}

We examine two settings regarding the strengths of players based on Elo points \citep{Elo1978}. While both setups are synthetic since the Elo ratings are assigned according to a linear pattern, the first is realistic and mimics the distribution of team strengths in a real-world chess tournament. The second is simplified to help measure ranking accuracy.

\begin{table}[t!]
\caption{Input data of the simulations}
\label{Table1}
\centering
\rowcolors{1}{}{gray!20}
\begin{tabularx}{\textwidth}{lCc} \toprule
                           & Setup A                         & Setup B  \\ \bottomrule 
Number of teams            & 32                              & 16  \\
Players per team     & 4                               & 4 \\
Elo rating of the best player  & 2750                            & 2750 \\
Within team Elo decrease & 40 for each player & None \\ 
Across team Elo decrease & 10 for each team   & 100 from 1st to 2nd; none after \\ \toprule
\end{tabularx}
\end{table}

Table~\ref{Table1} summarises the basic information about both setups. The 2025 European Team Chess Championship Open tournament (\url{https://s2.chess-results.com/tnr1263013.aspx}) has motivated Setup A. There were 40 participating teams. If the teams and the players within each team are ordered by their Elo ratings, the average rating difference between consecutive teams is approximately 13 points, while the average difference between consecutive players within the same team is around 43 points.
The highest-ranked team was Germany with an average Elo rating of 2681, and the second highest-ranked team was the Netherlands, with an average Elo rating of 2675; a difference of 6 points. For comparison, the average Elo of the fourth and fifth highest-ranked teams, Azerbaijan and England, differed by 21 Elo points.
For the individual players of the Dutch team, the average difference between the consecutive players is 39 points. Again, these differences vary. The Elo difference between the best and second-best players is 87 points, whereas the difference between the second- and third-best players is only 25 points.

The European Team Chess Championship is a fairly typical tournament, and Setup A models it in a stylized manner by rounding both average differences to the nearest multiple of ten and ignoring their variation. The advantage of using synthetic rather than real data resides in the transparent structure of expected winning probabilities. The strongest team has \emph{approximately} the same (expected) advantage over the second strongest team as the 19th-ranked team has over the 20th-ranked team, and likewise for any pair of consecutively ranked teams.
Even though a 100-point rating difference implies the same expected score in the Elo rating system, this does not hold for the estimated model of \citet{Milvang2015} used in the following (see Section~\ref{Sec32}). However, the differences are small.

Besides the variation in Elo differences, Setup A differs from the real tournament in two further aspects. First, the number of teams is fixed at 32. Second, the number of players in any team is four, but most teams enter the tournament with five players. They are listed in a fixed board order, which cannot be modified during the tournament, but the reserve player is allowed to play as a substitute on any board.

In Setup B, all players of the first team have an Elo rating of 2750, while all players of  the other teams have an Elo rating of 2650. Consequently, the strong team has a 71.5\% chance of winning the match against any other team, an 18.4\% chance of playing a draw, and a 10.1\% chance of losing. Thus, the first team is the clear favourite to win the tournament in Setup B.
As we will see in Section~\ref{Sec4}, if more rounds are played, then the probability that the first team is the only one with consistently strong performance increases, since in a match between two weak teams, they have roughly a 35\% chance of winning.
In the simple Setup B, the ranking of the teams is irrelevant except for the position of the first team, which simplifies the quantification of efficacy.

\subsection{The pairing method and the game prediction model} \label{Sec32}

For pairing the teams in each round of the tournament, we use an integer linear program that takes into account the current standings and the previous matchups. The objective is to pair teams with similar scores, subject to the constraint that no pair of teams can meet more than once. Hence, the team with the highest score is paired with the team with the second-highest score whenever possible. If this is infeasible, it is paired with the third-highest scoring team, and so on. Once the first pair has been determined, the procedure is repeated for the remaining teams.
This pairing rule is implemented by minimising the sum of squared score differences within pairs.

Formally, let $G = (V,E)$ be the tournament graph, where the nodes (edges) in set $V$ ($E$) represent the teams (matches). Let $F$ denote the set of feasible edges, corresponding to pairs of teams that have not played against each other in the earlier rounds. Denote by $s_u$ the current score of team $u\in V$.
Then the optimal solution to the following integer program gives the pairing in the next round:
\begin{align*}
\max \quad 
& \sum_{(u,v)\in F}
    \left(s_u-s_v\right)^2 x_{uv}\\[0.5em]
\text{s.t.}\quad
& \sum_{(u,v)\in F} x_{uv}=1,
    && \forall u\in V,\\
& x_{uv}\in\{0,1\},
    && \forall (u,v)\in F.
\end{align*}
If multiple optimal solutions exist, we use the solution returned by the solver in order to minimise running time. The LP was implemented in Python using the PuLP package, with COIN-OR Branch and Cut (CBC) as the default solver.

Pairing, including in the first round, is based exclusively on the current scores of the teams, which may be measured by board points, match points, or a combination of the two. Tie-breaking is not specified explicitly; the algorithm only checks whether the matching is valid, meaning that no two teams are paired more than once during the tournament, and whether the objective function is optimal. Seeding is ignored, although it may improve tournament efficacy \citep{SziklaiBiroCsato2022}.

For winning probabilities, we use the empirical distribution of \citet{Milvang2015}, similar to \citet{SauerCsehLenzner2024}.
It estimates win, draw, and loss probabilities from historical data, taking into account colour allocation since white advantage is well-documented in chess \citep{Henery1992, Fecker2024}. The model fits piecewise quadratic functions based on both the rating difference and the average rating of the two players.  Consequently, unlike in the standard Elo method, the predicted probabilities depend slightly on the absolute rating levels, as well as on the rating difference. In particular, the probability of a draw increases with the mean strength of the players.

\subsection{Tournament metrics} \label{Sec33}

Four alternative measures are used to quantify \emph{efficacy}:
\begin{itemize}
\item 
The average rank of the winner;
\item
Sum of Ranking Differences (SRD) computed from the true ranking and the final ranking of the teams;
\item 
The probability that the strongest team wins;
\item 
The probability that the strongest team finishes in the top three.
\end{itemize}
Sum of Ranking Differences is a comparative statistical method; see, for example, \cite{West2018} and \citet{GereSzakalHeberger2022}. SRD measures the distance between two rankings in $L^1$ norm. It ranges from 0 to 1; 0 indicates perfect agreement between two rankings, while 1 corresponds to complete disagreement, attained when one ranking is the reverse of the other. The distribution is not centred on 0.5 since the mean distance between a given ranking and a randomly generated ranking is approximately 0.66 \citep{SziklaiGereHebergerStaudacher2025}.

The average rank of the winner and SRD are only considered in Setup A because the 15 weak teams are identical in Setup B.
In the case of the first two measures, a lower value is preferred, but a higher value is better for the probabilities. Therefore, a reversed scale will be used for the average rank of the winner and SRD in the following.

Our \emph{competitiveness} metric is the probability of a \emph{non}-stakeless last round: the advantage of the team with the highest score after the penultimate round does not exceed the maximum number of points to be scored in one round. A higher value, ideally close to 1, is preferred.

To determine the final ranking, we consider different scoring rules.
The number of board points was used in the Chess Olympiad until 2006: one point for a win, half points for a draw, zero points for a loss, summed over the four boards. The maximum number of board points for a team in one round is four.
The number of match points has been used for ranking in the Chess Olympiad since 2008: two points for a win (at least 2.5 board points), one point for a draw (2-2 board points for the two teams, respectively), zero points for a loss (at most 1.5 board points). The maximum number of match points for a team in one round is two.
The 3-1-0 system, which is common in association football, is used in some prestigious individual chess competitions, such as the Bilbao Chess Masters Final and the London Chess Classic. We apply this rule to match points: three points for a win, one point for a draw, zero points for a loss. Then the maximum number of points for a team in one round is three.

As a robustness check, we also investigate whether combining board and game points can improve the trade-off between efficacy and competitiveness. For this purpose, the sum of board and match points is used, both with the 2-1-0 and the 3-1-0 system regarding match points. The maximum number of points for a team in one round is six and seven, respectively. 

Ties in the number of scores are broken by the traditional Sonneborn--Berger method without the lowest result: a team receives the match points of all teams it defeated and half the match points of all drawn opponents, but the lowest of these values is omitted.

\subsection{Implementation} \label{Sec34}

The number of rounds is fixed for any tournament and varies between 3--12 in Setup A (32 teams) and between 3--9 in Setup B (16 teams).
A round starts by pairing the teams according to their current match points, see Section~\ref{Sec32}. For each pair of teams, players are matched together according to their rating. Colours on board one are decided by play history. If the two lead players played with different colours in the previous round, their colours are reversed in the current round. If they had the same colour in the previous round, the player who had black more often is assigned white. Otherwise, the colour assignment is determined randomly.
The remaining players are assigned to the colours in an alternating manner. Thus, two-two players of every team play with the black and the white pieces, respectively, in each match.

The simulation was performed on a laptop equipped with an AMD Ryzen 9 8940HX processor and 32 GB of RAM, running Windows 11 Pro. For each tournament, 40 thousand simulation runs were performed to ensure convergence. Depending on the setup, the simulation took approximately 3--4 days to complete.

\section{Results} \label{Sec4}

\input{Figure1_efficacy_stakeless_tradeoff_average_rank_SRD}

Figure~\ref{Fig1} presents the relationship between efficacy (the average rank of the winner and SRD) and a non-stakeless last round in the realistic Setup A under the three basic scoring rules. As expected, efficacy improves when more rounds are played \citep{LasekGagolewski2018}---but this is parallel with a decline in competitiveness. The last round never becomes stakeless if the number of rounds is three or four. However, the probability of an exciting last round drops rapidly when the ranking is based on match points, and the situation is even worse with the 3-1-0 system.

The efficacy-competitiveness frontier is fully determined by the rule based on board points, when the chance of a stakeless last round remains below 3\% with 32 teams and 12 rounds. Since we do not know of any Swiss-system chess tournament consisting of more than 11 rounds, the threat to attractiveness remains limited with this ranking system. However, if the ranking is based on match points, the probability of a stakeless last round exceeds 6\% even with seven rounds.
Our findings are insensitive to the choice of the efficacy measure; the trade-off curve is quite similar when using the average rank of the winner and the Sum of Ranking Differences.

\input{Figure2_efficacy_stakeless_tradeoff_best_team}

As Figure~\ref{Fig2} shows, the main patterns are not affected by focusing on the strongest team or by considering a 16-team tournament with a clear favourite (Setup B). Ranking based on board points determines the Pareto frontier, and awarding three points for winning a match remains suboptimal. Setup B reveals that selecting the strong team as the winner is increasingly challenging if one wants to achieve a higher accuracy: for a mere improvement of 4 percentage points from 67\% to 71\%, the chance of an uninteresting last round grows from 3.3\% to more than 5.6\%. Organising a higher number of rounds has diminishing returns in terms of ranking accuracy, while its price in attractiveness is worryingly high.

\input{Figure3_efficacy_stakeless_all_ranking_systems}

Design choices such as using match points instead of board points or awarding extra points for match victories may serve valid objectives, but, according to Figures~\ref{Fig1} and \ref{Fig2}, they come at the cost of competitiveness if efficacy is to be maintained. In this respect, hybrid scoring systems are worth exploring, as they preserve the incentives created by these design choices to a great extent, while making it possible to move closer to the Pareto frontier.

Figure~\ref{Fig3} examines scoring systems based on the sum of board points and match points. Combining the two improves performance compared to using only match points (where a win is worth either two or three points); however, there remains a substantial gap from the Pareto frontier.


\section{Concluding remarks} \label{Sec5}

The current paper explores a previously unaddressed issue in tournament design: the trade-off between two key criteria, the accuracy of the final ranking and the competitiveness of the last round. To that end, we have examined the Swiss system, which ensures the best possible ranking for a given number of matches \citep{SziklaiBiroCsato2022}. The Swiss system is also widely used in chess and esports, while the number of rounds provide a crucial flexibility for decision-makers.

Our results uncover the need to balance the quality of the ranking and suspense in the last round: the former can be improved only by playing a higher number of rounds, which increases the likelihood that the winner will be determined early. This trade-off also depends on the scoring rule; according to the simulations, ranking is Swiss-system chess team tournaments should be based on the number of board points rather than the number of match points, while combining board and match points is preferred to using match points. These findings are remarkably robust to the choice of efficacy measure (average rank of the winner or SRD) and the assumed strength distribution (Setup A or B).

Therefore, the decision to replace board points with match points from the 2008 Chess Olympiad has been a questionable step. While the reform may have strengthened the team-competition nature of the tournament, ranking based on the number of match points seems to be suboptimal with respect to the crucial efficacy-competitiveness trade-off. Tournament organisers are encouraged to reconsider the scoring rule in light of our results.

A substantial scope remains for future research.
The pairing mechanism is a key feature of the Swiss system, and alternative pairing algorithms could enhance its performance. In particular, pairing the leading players against each other helps win the competition early by defeating the direct opponents. For the sake of attractiveness, however, it would be better to see the match between the strongest players towards the end of the tournament.
The second promising direction is offered by the analysis of incomplete round-robin tournaments. In contrast to the Swiss system, these competitions use a static schedule; the set of matches to be played is determined before the tournament starts. This difference might have a significant impact on the tournament metrics considered here.
Finally, the effect of awarding an extra point for team victory when teams consist of several players playing separate games needs to be investigated by empirical studies.

\section*{Acknowledgements}
\addcontentsline{toc}{section}{Acknowledgements}
\noindent
The research was supported by the National Research, Development and Innovation Office under Grants FK 145838, K 146320, Advanced 152220, and by the J\'anos Bolyai Research Scholarship of the Hungarian Academy of Sciences.

\bibliographystyle{apalike}
\bibliography{All_references}

\end{document}

%% file: Figure1_efficacy_stakeless_tradeoff_average_rank_SRD.tex
\begin{figure}[t!]
\centering

\begin{tikzpicture}
\begin{axis}[
name = axis1,
title = {Setup A, Average rank of the winner},
title style = {font=\small},
xlabel = Average rank of the winner (inverted scale),
x label style = {font=\small},
x tick label style={/pgf/number format/1000 sep=},
ylabel = Non-stakeless last round (\%),
y label style = {font=\small},
width = \textwidth,
height = 0.6\textwidth,
ymajorgrids = true,
x dir = reverse,
ymax = 102,
legend style = {font=\small,at={(-0.07,-0.15)},anchor=north west,legend columns=5},
legend entries = {Board points$\qquad$,Match points$\qquad$,Match points with three points for a win}
] 
\addplot [red, thick, dashed, mark=pentagon, mark size=2pt, mark options=solid] coordinates {
(4.33385,100)
(3.898525,100)
(3.64965,99.9725)
(3.43695,99.88)
(3.2934,99.5975)
(3.161825,99.3875)
(3.0828,99.035)
(2.981725,98.54)
(2.895375,97.94)
(2.814375,97.1875)
};
\addplot [ForestGreen, thick, mark=o, mark size=2pt, mark options=solid] coordinates {
(4.6806,100)
(4.3211,100)
(3.92115,100)
(3.724575,99.4975)
(3.53505,93.385)
(3.4296,90.4975)
(3.3175,88.865)
(3.218325,86.3425)
(3.12215,84.0525)
(3.055375,81.775)
};
\addplot [black, thick, dotted, mark=star, mark size=2.5pt, mark options=solid] coordinates {
(4.638325,100)
(4.314725,100)
(3.98,99.97)
(3.763375,96.6325)
(3.611125,91.0575)
(3.47715,87.2575)
(3.367875,83.8225)
(3.22915,81.52)
(3.125675,78.985)
(3.0686,76.085)
};
\end{axis}
\end{tikzpicture}

\vspace{0.5cm}
\begin{tikzpicture}
\begin{axis}[
name = axis1,
title = {Setup A, Sum of Ranking Differences (SRD)},
title style = {font=\small},
xlabel = Sum of Ranking Differences (SRD) (inverted scale),
x label style = {font=\small},
x tick label style={/pgf/number format/1000 sep=},
ylabel = Non-stakeless last round (\%),
y label style = {font=\small},
width = \textwidth,
height = 0.6\textwidth,
ymajorgrids = true,
x dir = reverse,
ymax = 102,
legend style = {font=\small,at={(-0.07,-0.15)},anchor=north west,legend columns=5},
legend entries = {Board points$\qquad$,Match points$\qquad$,Match points with three points for a win}
] 
\addplot [red, thick, dashed, mark=pentagon, mark size=2pt, mark options=solid] coordinates {
(0.33620654296875,100)
(0.30058330078125,100)
(0.276372265625,99.9725)
(0.2586435546875,99.88)
(0.24415419921875,99.5975)
(0.2331859375,99.3875)
(0.22399501953125,99.035)
(0.216085546875,98.54)
(0.2095248046875,97.94)
(0.20385419921875,97.1875)
};
\addplot [ForestGreen, thick, mark=o, mark size=2pt, mark options=solid] coordinates {
(0.35791328125,100)
(0.32544072265625,100)
(0.30132041015625,100)
(0.28122998046875,99.4975)
(0.264556640625,93.385)
(0.2515861328125,90.4975)
(0.2401662109375,88.865)
(0.230550390625,86.3425)
(0.2219044921875,84.0525)
(0.21480869140625,81.775)
};
\addplot [black, thick, dotted, mark=star, mark size=2.5pt, mark options=solid] coordinates {
(0.3651857421875,100)
(0.33480703125,100)
(0.30886376953125,99.97)
(0.28895859375,96.6325)
(0.2717255859375,91.0575)
(0.2578748046875,87.2575)
(0.2471517578125,83.8225)
(0.2371099609375,81.52)
(0.22832041015625,78.985)
(0.22145439453125,76.085)
};
\end{axis}
\end{tikzpicture}

\captionsetup{justification=centerfirst}
\caption{The efficacy-competitiveness trade-off: average rank of the winner and SRD \\ \vspace{0.2cm}
\footnotesize{\emph{Note}: The number of rounds increases from 3 to 12 (9) in Setup A (B) along the lines from left to right.}}
\label{Fig1}

\end{figure}


%% file: Figure2_efficacy_stakeless_tradeoff_best_team.tex
\begin{figure}[t!]
\centering

\begin{tikzpicture}
\begin{axis}[
name = axis1,
title = {Setup A, Strongest team wins},
title style = {font=\small},
xlabel = Probability of winning,
x label style = {font=\small},
x tick label style={/pgf/number format/1000 sep=},
ylabel = Non-stakeless last round (\%),
y label style = {font=\small},
width = 0.5\textwidth,
height = 0.4\textwidth,
ymajorgrids = true,
max space between ticks=40,
] 
\addplot [red, thick, dashed, mark=pentagon, mark size=2pt, mark options=solid] coordinates {
(0.145825,100)
(0.17815,100)
(0.202325,99.9725)
(0.220375,99.88)
(0.236225,99.5975)
(0.249,99.3875)
(0.2609,99.035)
(0.274325,98.54)
(0.278325,97.94)
(0.28965,97.1875)
};
\addplot [ForestGreen, thick, mark=o, mark size=2pt, mark options=solid] coordinates {
(0.212875,100)
(0.20045,100)
(0.216875,100)
(0.2353,99.4975)
(0.2476,93.385)
(0.252525,90.4975)
(0.262475,88.865)
(0.273725,86.3425)
(0.28195,84.0525)
(0.291075,81.775)
};
\addplot [black, thick, dotted, mark=star, mark size=2.5pt, mark options=solid] coordinates {
(0.218425,100)
(0.19415,100)
(0.21015,99.97)
(0.22435,96.6325)
(0.235925,91.0575)
(0.24895,87.2575)
(0.253625,83.8225)
(0.26295,81.52)
(0.2789,78.985)
(0.2828,76.085)
};
\end{axis}

\begin{axis}[
at = {(axis1.south east)},
xshift = 0.1\textwidth,
title = {Setup B, Strongest team wins},
title style = {font=\small},
xlabel = Probability of winning,
x label style = {font=\small},
x tick label style={/pgf/number format/1000 sep=},
ylabel = Non-stakeless last round (\%),
y label style = {font=\small},
width = 0.5\textwidth,
height = 0.4\textwidth,
ymajorgrids = true,
xmin = 0.29,
ymax = 102,
] 
\addplot [red, thick, dashed, mark=pentagon, mark size=2pt, mark options=solid] coordinates {
(0.406925,100)
(0.484725,99.9825)
(0.5517,99.7225)
(0.617075,98.71)
(0.670775,96.6925)
(0.711625,94.37)
(0.7476,90.9125)
};
\addplot [ForestGreen, thick, mark=o, mark size=2pt, mark options=solid] coordinates {
(0.3451,100)
(0.4227,100)
(0.4804,98.605)
(0.53405,88.07)
(0.582875,81.8775)
(0.618975,76.8875)
(0.66045,71.365)
};
\addplot [black, thick, dotted, mark=star, mark size=2.5pt, mark options=solid] coordinates {
(0.3404,100)
(0.4186,99.8225)
(0.4789,93.345)
(0.532875,86.0675)
(0.5804,77.945)
(0.62405,71.125)
(0.662625,66.0375)
};
\end{axis}
\end{tikzpicture}

\vspace{0.5cm}
\begin{tikzpicture}
\begin{axis}[
name = axis1,
title = {Setup A, Strongest team in top three},
title style = {font=\small},
xlabel = Probability of finishing in top three,
x label style = {font=\small},
x tick label style={/pgf/number format/1000 sep=},
ylabel = Non-stakeless last round (\%),
y label style = {font=\small},
width = 0.5\textwidth,
height = 0.4\textwidth,
ymajorgrids = true,
ymax = 102,
legend style = {font=\small,at={(-0.15,-0.25)},anchor=north west,legend columns=5},
legend entries = {Board points$\qquad$,Match points$\qquad$,Match points with three points for a win}
] 
\addplot [red, thick, dashed, mark=pentagon, mark size=2pt, mark options=solid] coordinates {
(0.40285,100)
(0.459425,100)
(0.49775,99.9725)
(0.528425,99.88)
(0.55955,99.5975)
(0.574225,99.3875)
(0.5942,99.035)
(0.6237,98.54)
(0.632725,97.94)
(0.657925,97.1875)
};
\addplot [ForestGreen, thick, mark=o, mark size=2pt, mark options=solid] coordinates {
(0.428175,100)
(0.46575,100)
(0.504375,100)
(0.527675,99.4975)
(0.55135,93.385)
(0.562625,90.4975)
(0.58095,88.865)
(0.603525,86.3425)
(0.617125,84.0525)
(0.637,81.775)
};
\addplot [black, thick, dotted, mark=star, mark size=2.5pt, mark options=solid] coordinates {
(0.431625,100)
(0.451,100)
(0.47895,99.97)
(0.5078,96.6325)
(0.52415,91.0575)
(0.5578,87.2575)
(0.569175,83.8225)
(0.5907,81.52)
(0.611675,78.985)
(0.627275,76.085)
};
\end{axis}

\begin{axis}[
at = {(axis1.south east)},
xshift = 0.1\textwidth,
title = {Setup B, Strongest team in top three},
title style = {font=\small},
xlabel = Probability of finishing in top three,
x label style = {font=\small},
x tick label style={/pgf/number format/1000 sep=},
ylabel = Non-stakeless last round (\%),
y label style = {font=\small},
width = 0.5\textwidth,
height = 0.4\textwidth,
ymajorgrids = true,
xmax = 0.945,
ymax = 102,
] 
\addplot [red, thick, dashed, mark=pentagon, mark size=2pt, mark options=solid] coordinates {
(0.67655,100)
(0.741925,99.9825)
(0.79725,99.7225)
(0.843675,98.71)
(0.876,96.6925)
(0.9049,94.37)
(0.923675,90.9125)
};
\addplot [ForestGreen, thick, mark=o, mark size=2pt, mark options=solid] coordinates {
(0.629675,100)
(0.69565,100)
(0.74855,98.605)
(0.79335,88.07)
(0.82785,81.8775)
(0.8583,76.8875)
(0.884125,71.365)
};
\addplot [black, thick, dotted, mark=star, mark size=2.5pt, mark options=solid] coordinates {
(0.62875,100)
(0.6946,99.8225)
(0.746625,93.345)
(0.7968,86.0675)
(0.8304,77.945)
(0.860275,71.125)
(0.8834,66.0375)
};
\end{axis}
\end{tikzpicture}

\captionsetup{justification=centerfirst}
\caption{The efficacy-competitiveness trade-off: Strongest team \\ \vspace{0.2cm}
\footnotesize{\emph{Note}: The number of rounds increases from 3 to 12 (9) in Setup A (B) along the lines from left to right.}}
\label{Fig2}

\end{figure}


%% file: Figure3_efficacy_stakeless_all_ranking_systems.tex
\begin{figure}[t!]
\centering

\begin{tikzpicture}
\begin{axis}[
name = axis1,
title = {Setup A, Average rank of the winner},
title style = {font=\small},
xlabel = Avg.\ rank of the winner (inv.~scale),
x label style = {font=\small},
x tick label style={/pgf/number format/1000 sep=},
ylabel = Non-stakeless last round (\%),
y label style = {font=\small},
width = 0.5\textwidth,
height = 0.4\textwidth,
ymajorgrids = true,
x dir = reverse,
ymax = 102,
max space between ticks=40,
] 
\addplot [red, thick, dashed, mark=pentagon, mark size=2pt, mark options=solid] coordinates {
(4.33385,100)
(3.898525,100)
(3.64965,99.9725)
(3.43695,99.88)
(3.2934,99.5975)
(3.161825,99.3875)
(3.0828,99.035)
(2.981725,98.54)
(2.895375,97.94)
(2.814375,97.1875)
};
\addplot [blue, thick, dashdotted, mark=square, mark size=2pt, mark options=solid] coordinates {
(4.6806,100)
(4.3211,100)
(3.92115,100)
(3.724575,99.4975)
(3.53505,93.385)
(3.4296,90.4975)
(3.3175,88.865)
(3.218325,86.3425)
(3.12215,84.0525)
(3.055375,81.775)
};
\addplot [ForestGreen, thick, mark=o, mark size=2pt, mark options=solid] coordinates {
(4.470175,100)
(3.9958,100)
(3.815575,99.975)
(3.6104,99.2425)
(3.44585,97.27)
(3.310825,94.8425)
(3.191475,92.53)
(3.091125,90.74)
(3.021825,88.44)
(2.933875,86.265)
};
\addplot [orange, thick, loosely dotted, mark=x, mark size=2.5pt, mark options=solid] coordinates {
(4.638325,100)
(4.314725,100)
(3.98,99.97)
(3.763375,96.6325)
(3.611125,91.0575)
(3.47715,87.2575)
(3.367875,83.8225)
(3.22915,81.52)
(3.125675,78.985)
(3.0686,76.085)
};
\addplot [black, thick, dotted, mark=star, mark size=2.5pt, mark options=solid] coordinates {
(4.76725,100)
(4.165475,100)
(3.898575,99.93)
(3.6563,97.925)
(3.511275,93.3625)
(3.36365,90.6625)
(3.2395,88.345)
(3.1622,86.245)
(3.0458,84.1175)
(2.970925,81.6)
};
\end{axis}

\begin{axis}[
at = {(axis1.south east)},
xshift = 0.1\textwidth,
title = {Setup A, Sum of Ranking Diff.},
title style = {font=\small},
xlabel = SRD (inv.~scale),
x label style = {font=\small},
x tick label style={/pgf/number format/1000 sep=},
ylabel = Non-stakeless last round (\%),
y label style = {font=\small},
width = 0.5\textwidth,
height = 0.4\textwidth,
ymajorgrids = true,
x dir = reverse,
ymax = 102,
] 
\addplot [red, thick, dashed, mark=pentagon, mark size=2pt, mark options=solid] coordinates {
(0.33620654296875,100)
(0.30058330078125,100)
(0.276372265625,99.9725)
(0.2586435546875,99.88)
(0.24415419921875,99.5975)
(0.2331859375,99.3875)
(0.22399501953125,99.035)
(0.216085546875,98.54)
(0.2095248046875,97.94)
(0.20385419921875,97.1875)
};
\addplot [blue, thick, dashdotted, mark=square, mark size=2pt, mark options=solid] coordinates {
(0.35791328125,100)
(0.32544072265625,100)
(0.30132041015625,100)
(0.28122998046875,99.4975)
(0.264556640625,93.385)
(0.2515861328125,90.4975)
(0.2401662109375,88.865)
(0.230550390625,86.3425)
(0.2219044921875,84.0525)
(0.21480869140625,81.775)
};
\addplot [ForestGreen, thick, mark=o, mark size=2pt, mark options=solid] coordinates {
(0.36242509765625,100)
(0.329269921875,100)
(0.302929296875,99.975)
(0.2827583984375,99.2425)
(0.2667169921875,97.27)
(0.25357080078125,94.8425)
(0.24269521484375,92.53)
(0.23426533203125,90.74)
(0.22601552734375,88.44)
(0.219444921875,86.265)
};
\addplot [orange, thick, loosely dotted, mark=x, mark size=2.5pt, mark options=solid] coordinates {
(0.3651857421875,100)
(0.33480703125,100)
(0.30886376953125,99.97)
(0.28895859375,96.6325)
(0.2717255859375,91.0575)
(0.2578748046875,87.2575)
(0.2471517578125,83.8225)
(0.2371099609375,81.52)
(0.22832041015625,78.985)
(0.22145439453125,76.085)
};
\addplot [black, thick, dotted, mark=star, mark size=2.5pt, mark options=solid] coordinates {
(0.361138671875,100)
(0.32878349609375,100)
(0.3032341796875,99.93)
(0.283721875,97.925)
(0.26736142578125,93.3625)
(0.2553923828125,90.6625)
(0.2447345703125,88.345)
(0.23596142578125,86.245)
(0.22835234375,84.1175)
(0.22149130859375,81.6)
};
\end{axis}
\end{tikzpicture}

\vspace{0.5cm}
\begin{tikzpicture}
\begin{axis}[
name = axis1,
title = {Setup A, Strongest team wins},
title style = {font=\small},
xlabel = Probability of winning,
x label style = {font=\small},
x tick label style={/pgf/number format/1000 sep=},
ylabel = Non-stakeless last round (\%),
y label style = {font=\small},
width = 0.5\textwidth,
height = 0.4\textwidth,
ymajorgrids = true,
ymax = 102,
legend style = {font=\small,at={(0.2,-0.25)},anchor=north west,legend columns=1},
legend entries = {Board points$\qquad \qquad \qquad \qquad \qquad \qquad \qquad \qquad \qquad \qquad \,\,$,Match points$\qquad \qquad \qquad \qquad \qquad \qquad \qquad \qquad \qquad \qquad \,$,Board points plus match points$\qquad \qquad \qquad \qquad \qquad \qquad \,$,Match points with three points for a win$\qquad \qquad \qquad \qquad$,Board points plus match points with three points for a win}
] 
\addplot [red, thick, dashed, mark=pentagon, mark size=2pt, mark options=solid] coordinates {
(0.145825,100)
(0.17815,100)
(0.202325,99.9725)
(0.220375,99.88)
(0.236225,99.5975)
(0.249,99.3875)
(0.2609,99.035)
(0.274325,98.54)
(0.278325,97.94)
(0.28965,97.1875)
};
\addplot [blue, thick, dashdotted, mark=square, mark size=2pt, mark options=solid] coordinates {
(0.212875,100)
(0.20045,100)
(0.216875,100)
(0.2353,99.4975)
(0.2476,93.385)
(0.252525,90.4975)
(0.262475,88.865)
(0.273725,86.3425)
(0.28195,84.0525)
(0.291075,81.775)
};
\addplot [ForestGreen, thick, mark=o, mark size=2pt, mark options=solid] coordinates {
(0.147,100)
(0.180975,100)
(0.2071,99.975)
(0.223275,99.2425)
(0.23625,97.27)
(0.248475,94.8425)
(0.261025,92.53)
(0.2711,90.74)
(0.277975,88.44)
(0.285275,86.265)
};
\addplot [orange, thick, loosely dotted, mark=x, mark size=2.5pt, mark options=solid] coordinates {
(0.218425,100)
(0.19415,100)
(0.21015,99.97)
(0.22435,96.6325)
(0.235925,91.0575)
(0.24895,87.2575)
(0.253625,83.8225)
(0.26295,81.52)
(0.2789,78.985)
(0.2828,76.085)
};
\addplot [black, thick, dotted, mark=star, mark size=2.5pt, mark options=solid] coordinates {
(0.142925,100)
(0.171775,100)
(0.201125,99.93)
(0.22025,97.925)
(0.2327,93.3625)
(0.245275,90.6625)
(0.25715,88.345)
(0.263775,86.245)
(0.2778,84.1175)
(0.2884,81.6)
};
\end{axis}

\begin{axis}[
at = {(axis1.south east)},
xshift = 0.1\textwidth,
title = {Setup A, Strongest team in top three},
title style = {font=\small},
xlabel = Probability of finishing in top three,
x label style = {font=\small},
x tick label style={/pgf/number format/1000 sep=},
ylabel = Non-stakeless last round (\%),
y label style = {font=\small},
width = 0.5\textwidth,
height = 0.4\textwidth,
ymajorgrids = true,
ymax = 102,
] 
\addplot [red, thick, dashed, mark=pentagon, mark size=2pt, mark options=solid] coordinates {
(0.40285,100)
(0.459425,100)
(0.49775,99.9725)
(0.528425,99.88)
(0.55955,99.5975)
(0.574225,99.3875)
(0.5942,99.035)
(0.6237,98.54)
(0.632725,97.94)
(0.657925,97.1875)
};
\addplot [blue, thick, dashdotted, mark=square, mark size=2pt, mark options=solid] coordinates {
(0.428175,100)
(0.46575,100)
(0.504375,100)
(0.527675,99.4975)
(0.55135,93.385)
(0.562625,90.4975)
(0.58095,88.865)
(0.603525,86.3425)
(0.617125,84.0525)
(0.637,81.775)
};
\addplot [ForestGreen, thick, mark=o, mark size=2pt, mark options=solid] coordinates {
(0.3981,100)
(0.443275,100)
(0.487775,99.975)
(0.5186,99.2425)
(0.537775,97.27)
(0.563925,94.8425)
(0.588925,92.53)
(0.60325,90.74)
(0.626875,88.44)
(0.640625,86.265)
};
\addplot [orange, thick, loosely dotted, mark=x, mark size=2.5pt, mark options=solid] coordinates {
(0.431625,100)
(0.451,100)
(0.47895,99.97)
(0.5078,96.6325)
(0.52415,91.0575)
(0.5578,87.2575)
(0.569175,83.8225)
(0.5907,81.52)
(0.611675,78.985)
(0.627275,76.085)
};
\addplot [black, thick, dotted, mark=star, mark size=2.5pt, mark options=solid] coordinates {
(0.383575,100)
(0.434175,100)
(0.472175,99.93)
(0.5086,97.925)
(0.536475,93.3625)
(0.559675,90.6625)
(0.5841,88.345)
(0.599925,86.245)
(0.620675,84.1175)
(0.6364,81.6)
};
\end{axis}
\end{tikzpicture}

\captionsetup{justification=centerfirst}
\caption{The efficacy-competitiveness trade-off: All ranking systems \\ \vspace{0.2cm}
\footnotesize{\emph{Note}: The number of rounds increases from 3 to 12 along the lines from left to right.}}
\label{Fig3}

\end{figure}
